\documentclass{aa}
\usepackage{graphicx}
\usepackage{txfonts}
\usepackage{natbib}
\usepackage{longtable}
\usepackage{hyperref} 
\usepackage{dsfont}
\hypersetup{colorlinks=true,linkcolor=blue,citecolor=blue,urlcolor=blue}
\usepackage{orcidlink}
\usepackage{lscape}
\usepackage{tabularray}
\usepackage{placeins}

\begin{document}





\title{Cosmic Type Ia SN rate and constraints on SN Ia progenitors}

\author { P. A. Palicio \orcidlink{0000-0002-7432-8709}  \inst{1} \thanks{email to: pedro.alonso-palicio@oca.eu} \and F. Matteucci \orcidlink{0000-0001-7067-2302} \inst{2,3,4}\and M. Della Valle \orcidlink{0000-0003-3142-5020} \inst{5,6}  \and  E. Spitoni \orcidlink{0000-0001-9715-5727}\inst{3} }
\institute{Universit\'e C\^ote d'Azur, Observatoire de la C\^ote d'Azur, CNRS, Laboratoire Lagrange, Bd de l'Observatoire,  CS 34229, 06304 Nice cedex 4, France  \and Dipartimento di Fisica, Sezione di Astronomia,
  Universit\`a di Trieste, Via G.~B. Tiepolo 11, I-34143 Trieste,
  Italy
\and I.N.A.F. Osservatorio Astronomico di Trieste, via G.B. Tiepolo
 11, 34131, Trieste, Italy 
\and I.N.F.N. Sezione di Trieste, via Valerio 2, 34134 Trieste, Italy
\and INAF - Osservatorio Astronomico di Capodimonte, Salita Moiariello 16, I-80131, Napoli, Italy 
\and ICRANet, Piazza della Repubblica 10, I-65122 Pescara, Italy
 }

 \date{Received xxxx / Accepted xxxx}

\abstract {
Type Ia supernovae play a key role in the evolution of galaxies by polluting the interstellar medium with a fraction of iron peak elements larger than that released in the core collapse supernova events. Their light-curve, moreover, is widely used in cosmological studies as it constitutes a reliable distance indicator at extra-galactic scales. Among the mechanisms proposed to explain the Type Ia SNe, the single and double degenerate channels are thought to be the dominant ones, which imply a different distribution of time delays between the progenitor formation and the explosion. }{In this paper, we aim at determining the dominant mechanism by comparing a {compilation of Type Ia SN rates with those computed from various cosmic star formation histories coupled with different delay time distribution functions, and evaluating the relative contributions of both channels}.}{By using a least-squares fitting procedure, we model the observations of Type Ia SN rates assuming different combinations of three recent cosmic star formation rates and seven delay time distributions. The goodness of these fits are statistically quantified by the $\chi^2$ test. }{For two of the three cosmic star formation rates, the single degenerate scenario provides the most accurate explanation for the observations,{while a combination of 34\% single degenerate and 66\% double degenerate delay time distributions is more plausible for the remaining tested cosmic star formation rates.}} {Though dependent on the assumed cosmic star formation rate, we find arguments in favor of the single degenerate model. From the theoretic point of view, {at least the $\sim$~34\% of the Type Ia SN} must have been produced through the single degenerate channel to account for the observations. The wide double degenerate scenario mechanism slightly under-predicts the observations at redshift $z\gtrsim 1$, unless the cosmic SFR flattens in that regime. On the contrary, although the purely close double degenerate scenario can be ruled out, we cannot rule out a mixed scenario with single and double degenerate progenitors.}


\titlerunning{Constraints on SN Ia progenitors}

\authorrunning{Palicio et al.}

\maketitle

\section{Introduction}
\label{Sect_intro}
\par The study of Type Ia supernovae (SNe) has significant implications for the cosmology and galactic astronomy, allowing us to construct the Hubble diagram at low and high redshifts to constrain some cosmological parameters e.g. \citet{Hamuy1995, Perlmutter1998, Perlmutter1999a, Perlmutter1999b, Riess1998, Sullivan2011, Suzuki2012, Ganeshalingam2013, Betoule2014, Rest2014, Khetan2021} and, as major producers of iron \citep{greggio1983, matteucci1986}, to model the chemical evolution of galaxies among other applications \citep[see review of][]{Lapuente14}. Given this variety of applications, understanding the nature of their progenitors is crucial. In the past years, many Type Ia SN progenitor models have been proposed both theoretically and empirically, in which the most common scenarios are the double degenerate (DD) and single degenerate (SD) channels (see \citealt{LivioMazzali18} for a recent review), {although other alternatives have emerged for accounting for some observations \citep[see review of ][and references therein]{Soker2024}}. The SD scenario was originally proposed  by \citet{Whelan1973} and consists of a system formed by a C-O white dwarf (WD) plus a normal star that, evolving into a red giant, transfers material over the WD \citep{Wheeler1971, Nomoto1982a, Nomoto1982b}. The WD then explodes when its mass reaches the Chandrasekhar mass limit ($\sim$~$1.44 \rm M_{\odot}$). In the DD scenario, originally proposed by \citet[][but see also \citealt{KatoHachisu12, KatoHachisu15}]{Iben1994}, two WDs in a binary system merge after emission of gravitational waves and explode when reaching the Chandrasekhar mass \citep{Tutukov1976, Tutukov1979, Webbink1984}. The relative contribution of the two main Type Ia SN mechanisms has been historically controversial, with no fully satisfying scenario \citep{Livio2000, Greggio08,Valiante2009, matteucci2009,bonaparte2013, Maoz2014review, Ruiz-Lapuente2019}. Originally, the preferred explosion mechanism was the C-deflagration of Chandrasekhar WD, but later other possible mechanisms were proposed in order to account for a variety of Type Ia SN types, requiring a different exploding mass and involving also He and C-detonation besides C-deflagration (see \citealt{Hillebrandt2013} for a review). The discovered variety of Type Ia SNe {required} different amounts of $\rm  ^{56}Ni$ to power the light curve, for a while created uncertainty on the possibility of using the Type Ia SNe as standard candles. However, \citet{Phillips1993} discovered a relation between the magnitude at maximum of Type Ia SNe and that after 15 days that allows us to still use these SNe as standard candles.
\par The delay between the progenitor formation and its ultimate explosion is statistically determined by the delay-time distribution (DTD), initially introduced by \citet[][see their figure 2]{Madau98} for Type Ia SN, found its first publication in its complete analytical formulation through the work of \citet{greggio2005}. Such a DTD can have an analytical form and can describe either the classical SD and DD models or other empirical laws derived from direct observations of Type Ia SNe.
\par In this work, we will treat neither the chemical evolution nor the explosion mechanisms, but the progenitors of these systems. More precisely, we aim at computing the cosmic Type Ia SN rate ($\rm Ne\ yr^{-1}Mpc^{-3}$) by adopting observed cosmic star formation rates (CSFRs) convolved with a variety of delay time distribution functions associated with different progenitor models. We test several suggested CSFRs, such as that of \citet{madau14}, \citet{harikane22} and \citet{jinkim23} and seven different DTD functions: the DTD of \citet{matteucci2001} for the SD model, the wide and close DTDs of \citet{greggio2005} for the DD model and the empirically derived ones proposed by \citet{totani2008}, \citet{mannucci2006}, \citet{Pritchet08} and \citet{strolger2005}. By comparing the predicted cosmic Type Ia rates against the observational data at different redshifts, we put constraints on the best DTD and, therefore, on the best progenitor model for Type Ia SNe. {In the previous years, \citet{bonaparte2013, Graur14, Rodney14, Strolger20} presented similar works adopting different CSFRs and DTDs, both theoretical and observed \citep{madau1998}}. Here we adopt the most recent data on Type Ia SN cosmic rate.
\par The paper is organized as follows: in Section \ref{Sect_DTD} we introduce the notion of Type Ia SN rate computed by means of various adopted DTDs functions, in Section \ref{Sect_CSFR} we present the best fits to the observed CSFRs together with the computation of the cosmic Type Ia SN rate. In Section  \ref{Sect_obstheo} we describe the observational data and a comparison between observations and theory. In Section \ref{Sect_results} we present and discuss the theoretical results compared to the observations at different redshifts.
Finally, in Section \ref{Sec_conclu} we draw some conclusions.
\section{The DTD functions}
\label{Sect_DTD}
The DTD function for SN Type Ia was first proposed by \citet{greggio2005}, and in the following we will adopt the same formalism as in that paper. 
The Type Ia SN rate, $R_{SNIa}$, can be written as:
\begin{equation}
    R_{SNIa}=k_{\alpha} \int_{\tau_i}^{min(t, \tau_x)}{A_{\rm Ia}(t-\tau) \psi(t-\tau) DTD(\tau)d\tau}
    \label{eq_rsnia}
\end{equation}
where $k_{\alpha}$ is the number of stars per unit mass in a stellar generation and is determined by the initial mass function (IMF) as:
\begin{equation}
k_{\alpha}=\frac{\int^{M_U}_{M_L}{\phi(m)dm}}{\int^{M_U}_{M_L}{m\cdot\phi(m)dm}}.
\end{equation}
The denominator of this expression is the normalization condition of the IMF and is equal to 1. 
Here we adopt the \citet{salpeter55} IMF, which is the same IMF adopted to derive the CSFRs that we will use to compute the cosmic Type Ia SN rate. 
The quantity $\phi(m)$ is the IMF and it is chosen to agree with that adopted in deriving the CSFR, as we will see in the following Section. The masses{ in the integration limits are }$M_L=0.1M_{\odot}$ and $M_U=100M_{\odot}$.
The quantity $\psi(t)$ is the star formation rate (SFR). {The quantity DTD(t) is the assumed delay time distribution function}. It is defined in a range of times ($\tau_i$, $\tau_x$), which represent the minimum and maximum explosion times for the SNe Ia. The DTD represents {an instantaneous burst of star formation} and therefore needs to be convolved with the SFR in order to obtain the SN rate.
The quantity $A_{\rm Ia}$ represents the fraction of systems exploding as Type Ia SNe relative to all the stars defined for the mass range of the IMF. {Although \(A_{\rm Ia}\) might vary with redshift due to a possible dependence on metallicity, we assume it to be constant as per usual practice \citep{matteucci2021}, since there is no consensus on a precise metallicity dependence. On one hand, if the metallicity of the accreted gas is $\lesssim -1$~dex, the wind from the white dwarf (WD) is too weak to trigger a Type Ia supernova explosion \citep{Kobayashi1998, Kobayashi2009}. On the other hand, the fraction of close binaries is higher in metal-poor populations \citep{ElBadryRix18, Moe2019}. Furthermore, the cosmic Type Ia SN rates computed in \citet{Kobayashi1998} for the scenarios with and without such metallicity dependence show small differences at redshift $z\lesssim 1.2$, where the majority of our sample lies (see Tables \ref{Table_binned_RIa} and \ref{Table_full_RIa}). Similarly, we assume that the shape of the DTD does not change with metallicity, although variations are observed in the single degenerate scenario compared to the double degenerate scenario \citep{Meng11, Meng12}, which is less affected by metallicity. }
\begin{figure}
\begin{centering}
\includegraphics[width=0.5\textwidth]{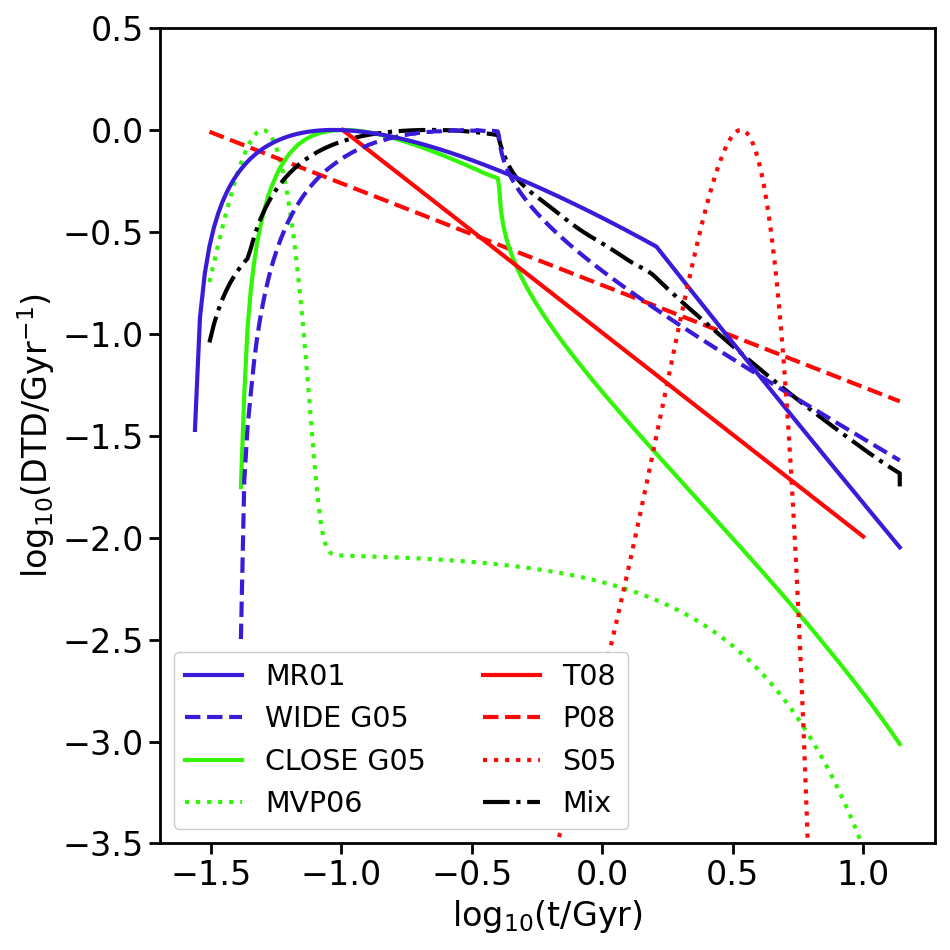}
\caption{The different DTD functions assumed in this work: {\citet[][solid blue curve]{matteucci2001}, \citet[][dotted green curve]{mannucci2006}, wide and close from \citet[][dashed blue and solid green curves, respectively]{greggio2005}, \citet[][solid red curve]{totani2008}, \citet[][dashed red curve]{Pritchet08}, and \citet[][dotted red curve]{strolger2005}}. The optimal combination of DTDs detailed in Sec. \ref{Sec_mixedDTDs} is denoted by the {dotted-dashed black line}. All the distributions are normalized by their maximum.}
\label{Fig_DTD}
\end{centering}
\end{figure}
In Figure \ref{Fig_DTD} we show the different DTD(t) functions adopted in this paper. As one can see in the \citet[][hereafter MVP06]{mannucci2006} the fraction of prompt Type Ia SNe (those exploding during the first 100 Myr from the beginning of star formation) is as large as 50\%, whereas in the \citet[][hereafter S05]{strolger2005} there are no prompt SNe-Ia Type. In all the other DTDs the fractions of prompt Type Ia SNe is below 20\%. {It is worth noting the power-law trend at late times of the \citet[][hereafter T08]{totani2008} DTD}, the one of \citet[][hereafter MR01]{matteucci2001} and these of \citet{greggio2005} for the wide and close double degenerate scenarios (hereafter G05Wide and G05Close, respectively).
\par {The T08 DTD has a form of $\sim t^{-1}$ and was proposed to match the Type Ia supernova DTDs inferred from a sample of old, faint variable objects detected in the Subaru/XMM-Newton Deep Survey \citep[SXDS,][]{Furusawa08} at $z \sim 0.4-1.2$. Similarly, \citet{Pritchet08} found that a power law with an index of -1/2 accounts for the Type Ia SN rates reported by \citet{Sullivan06} from the Supernova Legacy Survey \citep[SNLS,][]{Astier06}. Using a Bayesian maximum likelihood method, \citet{strolger2004} concluded that a Gaussian DTD better fits the redshift distribution of Type Ia supernovae inferred from the \textit{Hubble} Higher $z$ Supernova Search project \citep[HHZSS,][]{Schmidt1998} and the Great Observatories Origins Deep Survey \citep[GOODS,][]{Giavalisco2004}. On the contrary, \citet{mannucci2006} found evidence of a bimodal DTD, where prompt Type Ia SN explosions peak at approximately 50 Myr after progenitor formation, while the remaining events follow a more extended exponential distribution with a timescale of 3 Gyr. The analytic formulations of the DTDs proposed by \citet{matteucci2001} and \citet{greggio2005} for the SD and DD scenarios, respectively, are characterized by a plateau at early times followed by a power-law decreasing trend (see Fig. \ref{Fig_DTD}). Due to their complexity, we refer to their respective papers for detailed explanations.}
\section{The CSFR and the calculation of the cosmic Type Ia rate}
\label{Sect_CSFR}
The CSFR is the star formation rate in a unitary volume of the universe and contains the SFR of a mixture of galaxies. It depends upon the distribution of galaxies as a function of redshift. Several attempts to model the CSFR have appeared in the past years and they take into account precise cosmological scenarios of galaxy formation. For example, in \citet{gioannini2017}, the SFR of galaxies of different morphological types (ellipticals, spirals and irregulars) were computed by means of detailed chemical evolution models, then these histories of star formation were convolved with the number density of galaxies as a function of redshift. The galaxy number density was assumed either to be constant in time (Pure Luminosity Evolution) or to vary with redshift following the paradigm of the hierarchical galaxy formation.

\begin{figure}
\begin{centering}
\includegraphics[width=0.49\textwidth]{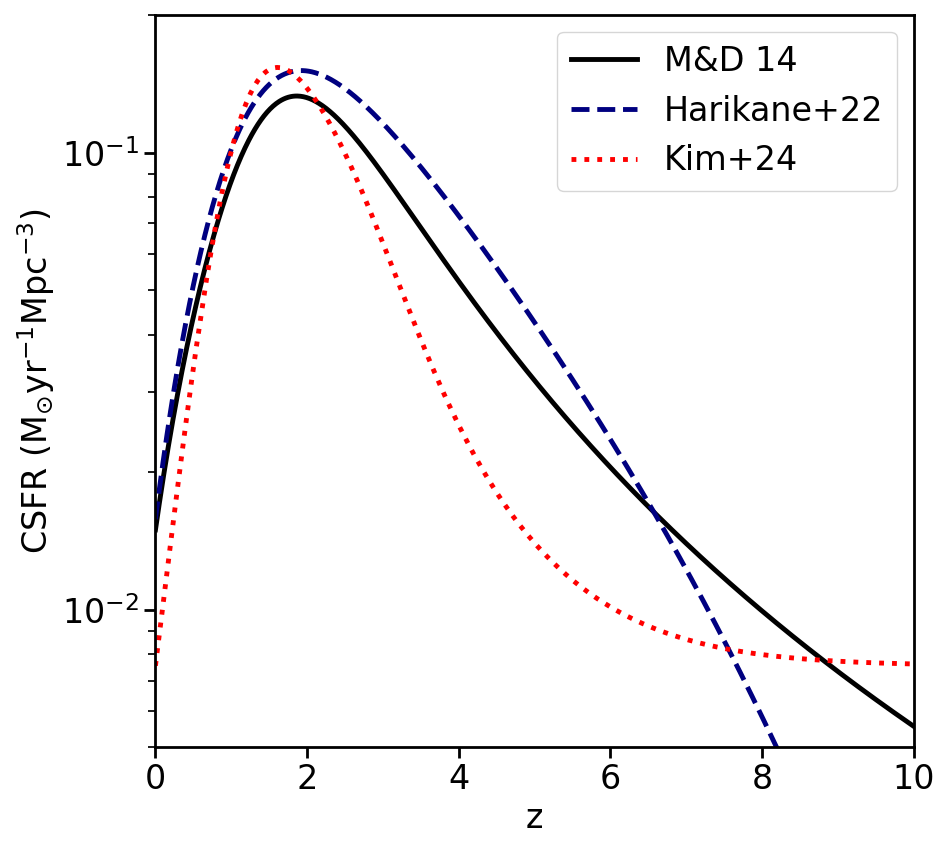}
\caption{The curves here represent the adopted CSFRs derived by the best fit to observations as a function of redshift. {The black solid line shows the CSFR of \citet{madau14}, the blue dashed line corresponds to the CSFR of \citet{harikane22},  while the red dotted line represents the CSFR of \citet{jinkim23}.}}
\label{Fig_CSFR}
\end{centering}
\end{figure}

Among the prescriptions for a hierarchical formation, the best number density evolution resulted an empirical one, as derived by \citet{pozzi2015}, which provided a theoretical CSFR in excellent agreement with the observationally derived CSFR by \citet{madau14}.
Here we will adopt only observationally derived CSFR (in units of M$_{\odot}$ yr$^{-1}$ Mpc$^{-3}$) and, in particular, the CSFR of \citet{madau14}, the one of \citet{harikane22} and that of \citet{jinkim23}. The best fits of these empirically derived CSFRs are shown in Figure \ref{Fig_CSFR}. As one can see, the CSFRs are very similar in the redshift range $z=3-0$, whereas they differ for $z>3$.

\par In order to compute the cosmic rate of SNe Ia ($R_{\rm IaCOSM}$), we adopt equation (\ref{eq_rsnia}), where the SFR is given by the CSFR. The cosmic Type a rate will be in units of SNe yr$^{-1}$ Mpc$^{-3}$.
In particular:

\begin{equation}
    \label{Eq_R1aCosmo}
    R_{\rm IaCOSM}= k_{\alpha} A\int_{\tau_i}^{min(t, \tau_x)} {\rm CSFR}(t-\tau) \, DTD(\tau) \, d\tau,
\end{equation}

with $k_{\alpha}$=2.846 M$_{\odot}^{-1}$ for a Salpeter IMF. On the contrary, the fraction $A$ constitutes a free parameter to be set by the optimal fit of the data for each assumed DTD (see Section \ref{Sect_results}), with typical values of $\sim10^{-4}-10^{-2}$ (see Table 1 of \citealt{bonaparte2013}).
\par {We estimate the error of $R_{\rm IaCOSM}$ from the reported errors for the CSFRs following the procedure described in \citet{Klein21}. However, for the CSFR of \citet{madau14}, since no uncertainties are provided, we perform 1,000 random realizations of their compilation of observational data and use the standard deviation of all the fits as the error.}
\par Once the $R_{\rm IaCOSM}$ is computed, we compare it to observational data referring to the cosmic Type Is SN rate. The SNe-Ia rates derived from observations are reported in Table \ref{Table_full_RIa} including all SNe-Ia rates calculated by various authors in different surveys over the past 25 years. In col. 1, we present the redshift; in col. 2, the SN rate; in col. 3 and 4, the associated errors, and in col. 5, the reference source. The SN-Ia rate is reported in units of $10^{-4}$ yr$^{-1}$ Mpc$^{-3}$ for $H_0=70$~km s$^{-1}$ Mpc$^{-1}$ \citep{Altavilla04, Dhawan18, Khetan2021}. The measurements exhibit relatively large error bars, and only the most ``extreme'' theoretical models might be excluded. Consequently, we have prepared a second table (Table \ref{Table_binned_RIa}) where we calculated the weighted averages of SN rates at similar redshift values (where possible). Table \ref{Table_binned_RIa} reports the average redshift with its dispersion in the first and second columns, respectively, the average SN rate in the third column, as well as its associated upper and lower error limits in the fourth and fifth columns, respectively. The use of average rates instead of single measurement enables a more effective constraining of the theoretical models to the data.
%
%
%

\begin{table}
\begin{center}
\begin{tabular}{ccccc}
\hline
  \hline
  z	&	$\Delta$z &          Rate 	&	$\rm \sigma +$	&	$\rm \sigma -$\\ 
    \hline
    
0.01	&	0	&	0.183	&	0.046	&	0.046	\\	 
0.02	&	0	&	0.265	&	0.055	&	0.054	\\
{0.024}	&	{0}	&	{0.228}	&	{0.020}	&	{0.020}	\\
0.0375	&	0	&	0.278	&	0.115	&	0.083	\\	 
0.073	&	0	&	0.242	&	0.043	&	0.034	\\	 
0.09	&	0	&	0.29	&	0.09	&	0.07	\\
0.102	&	0.005	&	0.2506	&	0.0281	&0.0286 \\
0.1425	&	0.0096	&	0.254	&	0.0392	&	0.0280\\
0.2	&	0	&	0.348	&	0.088	&	0.031		 \\
0.2525 &		0.005	&	0.3455	&	0.0794	&	0.0284\\
0.3	&	0	&	0.369	&	0.220	&	0.037\\
0.356	&	0.010	&	0.360	&	0.062	&	0.053\\
 0.407	&	0.011	&	0.505	&	0.284	&	0.140\\
0.452	&	0.0086	&	0.4296	&	0.057	&	0.053\\
0.55	&	0	&	0.52	&	0.064	&	0.065\\
0.644	&	0.0134	&	0.602	&	0.070	&	0.077\\
 0.747	&	0.0052	&	0.717	&	0.078	&	0.083\\
0.802	&	0.0035	&	0.921	&	0.164	&	0.152\\
0.843	&	0.011	&	0.674	&	0.088	&	0.092\\
0.947	&	0.0058	&	0.775	&	0.123	&	0.120\\
1.05	&	0	&	0.841	&	0.170	&	0.189\\
1.187	&	0.0277	&	0.857	&	0.180	&	0.147\\				
1.237	&	0.011	&	0.844	&	0.197	&	0.196 \\
1.558	&	0.028	&	0.389	&	0.286	&	0.106\\	
1.664	&	0.061	&	0.590	&	0.219	&	0.152\\
2.1	&	0	&	<1.07   &              0.0	&	0.0	\\	 
2.25	&	0	&	0.49 &		1.05	&	0.449\\	 
\hline 
 \end{tabular}
\caption{Weighted averages of SN rates (third column) calculated based on similar redshift values (first column), when possible. Bin size is indicated in the second column while the upper and lower errors are summarised in the fourth and fifth columns, respectively. Type Ia SN rates and their errors are in units of $\rm 10^{-4} yr^{-1} Mpc^{-3}$. }
\label{Table_binned_RIa}
\end{center}
\end{table}
 \section {Observations vs. Theory}
\label{Sect_obstheo}
\par The fundamental properties of Type Ia SNe derived directly from observations can be summarized as follows \citep{Livio2000}: i) Type Ia SNe are observed in both early and late Hubble types. ii) Their occurrence in early-type galaxies \citep{Turatto94}, characterized by their quenched star formation rate, suggests that SNe-Ia are caused by stars with masses smaller than $\rm 8 M_{\odot}$. iii) The absence of hydrogen and helium in the ejecta would indicate highly evolved progenitors. iv) The energy per unit mass is comparable to that obtained from the conversion of carbon and oxygen into iron, approximately $0.5\times (10^4 \rm{km/s})^2$. Based on these observations, the prevailing consensus suggests that Type Ia SNe stem from the thermonuclear disruption of mass-accreting WDs in binary systems, as outlined in the two scenarios discussed in Section \ref{Sect_intro}.
\par On the other hand, deep pre-explosion images obtained with the \textit{Hubble} Space Telescope of the site of the ``Branch normal'' SN-Ia 2011fe, that exploded in the nearby galaxy M101 \citep{Nugent2011,Li2011Natur}, did not reveal a suitable candidate progenitor. This led to the conclusion that the progenitor must have been fainter than previously considered candidates by a factor of 10-100, which consequently excludes luminous red giants as well as the vast majority of helium stars as potential mass donating companions. This symbiotic progenitor was also ruled out by the analysis of X-ray observations performed by \citet{Horesh12} and \citet{Meng16}, but other supersoft sources \citep{Lapuente14} are possible. On the contrary, \citet[][but see also \citealt{mccully22}]{mccully2014} reported the detection of a white dwarf supernova progenitor in the pre-explosion images of SN-Iax 2012Z \citep{Cenko12}, and proposed it was constituted by a helium star and a white dwarf. As noted by \citet{mccully2014}, once SN 2012Z has faded, new observations will either validate their hypothesis or reveal that this supernova was, in fact, the explosive end of a massive star. However, this proposition is not supported by the observations of \citet{mccully22}, given that the blue object is not only still present but also retains even greater luminosity. The mentioned pre-explosion observations of SN-Ia 2011fe suggest a scenario in which either another WD or the overflow from the Roche-lobe of a subgiant or main-sequence companion provides the material accreted by the exploding WD. Among these latter systems, Recurrent Novae (RNe) have been often considered suitable candidates (e.g. \citealt{KatoHachisu12}), although this suggestion was not supported by observations \citep{dellaValle1996, dellaValle2020,Shafter2015}. Both teams concluded that it is very unlikely that RNe could play a major role as Type Ia SN progenitors. Lastly, it is noteworthy that the radio observations of 27 Type Ia SNe conducted with the Very Large Array by \citet{Panagia2006} did not yield a single positive detection. This lack of a radio signal indicated an exceptionally low circumburst density, enabling the establishment of an upper limit on the mass loss rate of the secondary star of approximately $\dot{M} \sim 3\times 10^{-8}$~$\rm M_{\odot}/yr$. This rate is at least one order of magnitude smaller than the $\dot{M}$ expected from a red giant donor. Subsequently, \citet{Chomiuk2016} further constrained this upper limit to $5\times 10^{-9}$~$\rm M_{\odot}/yr$. 
\par At this point the question arises as to whether there is any possibility for the existence of the single degenerate scenario, or if it has been entirely excluded. Recently, \citet{Darnley2006, Darnley2019} presented compelling evidence supporting the idea that the progenitor system of the recurrent nova M31N 2008-12a, located in Andromeda galaxy and which undergoes eruptions on an annual basis, is a single degenerate system featuring a red giant donor (see also \citealt{Williams2016}). Additionally, \citet{KatoHachisu15} determined that the central WD in this system is of CO type and possesses a mass of approximately $\rm \sim 1.38 \rm M_{\odot}$. Notably, the WD is experiencing rapid mass growth, expelling only about 40\% of the accreted matter during each nova outburst cycle, equivalent to about $10^{-7}$~$\rm M_{\odot}$. Consequently, the Chandrasekhar limit should be achieved within approximately 100,000 years, potentially culminating in a Type Ia supernova eruption. In addition, we point out the existence of recent hydro-dynamical simulations by \citet{Starrfield2020} that suggest that carbon-oxygen WDs in the mass range of $\rm 0.6-1.35 M_{\odot}$, accreting material from companions at a rate of approximately $\rm 1.6 \times 10^{-10} M_\odot/yr$, can grow in mass up to 80\% of the accreted mass after each nova cycle. However, these results are preliminary and require further validation. 
\par {The detection of circumstellar material (CSM) has historically been consider a proof of the single degenerate mechanism. Comparing spectra of 2006X at different epochs, \citet{patat2007} found mean velocities in the CSM compatible with the winds of a red giant star. Given the lack of peculiarities in optical, UV and radio, the authors found no reason to consider 2006X an exceptional case. Similarly, \citet{dilday2012} found evidence for an WD+RG single degenerate scenario in the interaction of the CSM with the supernova ejecta of PTF 11kx, while \citet{Sternberg2011} considered a sample of 35 Type Ia SNe to conclude the 20-25\% of Type Ia SNe in spirals are produced through the SD channel. On the other hand, the absence of hydrogen in some Type Ia SNe observations is explained by the gain of angular momentum by the WD during the accretion phase and the subsequent contraction of the donor's envelope \citep{Justham2011}. The extra angular momentum allows the WD to exceed the Chandrasekhar limit without collapsing or exploding, thereby accreting more material from its companion's envelope. This results in significant mass loss and rapid contraction of the donor envelope, which reduces the cross-section of the companion and thus the interaction with the supernova shock when the WD eventually spins down and explodes. Furthermore, the associated delay between mass growth and explosion might allow the ejected material to diffuse within the interstellar medium \citep{Distefano2011}; while the remaining mass of the donor envelope might be below the observational detection limit \citep[$\lesssim 0.01 M_{\odot}$,][]{Leonard07}. According to \citet{Justham2011}, although this model is most effective for the WD+RG configuration, it may also describe other SD scenarios under exceptional circumstances.}
\par {If the accretion rate exceeds a critical value\footnote{This critical value corresponds to the maximum accretion rate that allows stable hydrogen burning on the surface of the white dwarf.}, a portion of the accreted material is not burned but instead accumulates around the white dwarf, forming a red giant-like object. This object fills the Roche Lobe and creates a common envelope, which prevents a supernova explosion. By considering the opacities from \citet{Iglesias87,Iglesias90,Iglesias91,Iglesias93, Rogers92}, \citet{Hachisu1996} suggested that the unprocessed material escapes from the system as an optically thick wind (OTW), stabilizing the mass transfer from $M_2=0.79 M_{WD}$ to $1.15 M_{WD}$ and aligning the Type Ia supernova birth rate with observations. However, the creation of this optically thick wind requires a minimum metallicity threshold \citep{Kobayashi1998, Meng2009}, which has not been observed \citep{Galbany16, Prieto2020}. Contrarily, Type Ia supernova events have been reported in low metallicity populations \citep{Frederiksen12}. To explain these observations, \citet{meng2017} proposed an alternative model in which the loss of unprocessed material is driven by common envelope winds (CEWs). This model does not depend on metallicity and allows the white dwarf's mass-growth rate to be higher than in the OTW model. Both the OTW and CEW models can explain the small number of supersoft X-ray sources (SSSs) observed in ellipticals and spirals compared to the predictions of the standard SD model \citep[][but see also \citealt{Hachisu1996} and Section 3.1.2 in \citealt{Maoz2014review}]{DiStefano10, Gilfanov10} through the absoption of the X-radiation by the optically thick and common envelope winds, respectively.}
\par {There is observational evidence suggesting multiple progenitor configurations for Type Ia SNe. \citet{Wang2013} discovered a bimodality in the expansion velocity of 123 "Branch normal" Type Ia SNe observed in external galaxies. High expansion velocity supernovae make up approximately one-third of the sample and are typically located closer to the galactic center, likely originating from a younger, higher metallicity population compared to those SNe within normal velocity peaks. Similarly, \citet{Rigault13} examined the relationship between the properties of Type Ia SNe and their formation environments. They found differences between passive evolving and star-forming environments in the color and stretch of supernovae.}
\par In the last decade, a third channel known as core-degenerate scenario has been proposed as alternative to the SD and DD mechanisms \citep{Kashi11, Ilkov12, Soker13IAU, Soker2024}. In this scenario, the C-O WD merges with the core of an AGB star, resulting in a more massive white dwarf whose fast rotation prevents Type Ia SN until it spins down. \citet{SokerPTF_2013, Soker14} invoked this mechanism to explain the PTF 11kx and SN 2011fe supernovae \footnote{More precisely, the PTF 11kx SN is explained by the violent prompt merger mechanism, a subcase of the CD scenario in which the Type Ia SN is produced shortly after the merger within the common envelope.}. However, it is estimated the CD mechanism accounts for less than 1\% of all the Type Ia SNe \citep{Meng12}.
\par Although our results favour the single degenerate scenario, it is plausible that both single degenerate and double degenerate scenarios contribute to Type Ia supernova events, but a revision of these models, as well as more exotic mechanisms \citep{Soker2024}, might be considered to account for the discrepancies with the observations mentioned above.
%
%
%
%
%
%
\section{Results and Discussion}
\label{Sect_results}
\subsection{Results with literature DTDs}
%
%
\begin{table*}
\centering
\begin{tblr}{
  cell{1}{2} = {c=3}{c},
  cell{1}{5} = {c=3}{c},
  cell{1}{8} = {c=3}{c},
  cell{2}{2} = {c},
  cell{2}{3} = {c},
  cell{2}{4} = {c},
  cell{2}{5} = {c},
  cell{2}{6} = {c},
  cell{2}{7} = {c},
  cell{2}{8} = {c},
  cell{2}{9} = {c},
  cell{2}{10} = {c},
  vline{3,6,9} = {1}{},
  vline{1-2,5,8,11} = {2-10}{},
  hline{1} = {2-10}{},
  hline{2-3,10} = {-}{},
  hline{2-3,11} = {-}{},
}
         &  M\&D14 CSFR       &          &       & H22 CSFR         &          &       & K24 CSFR         &          &       \\
DTD & $\chi^2/{dof}$ & $\hat{A}_{\rm Ia}(\times 10^{-4})$ & $1-p$ & $\chi^2/{dof}$ & $\hat{A}_{\rm Ia}(\times 10^{-4})$ & $1-p$ & $\chi^2/{dof}$ & $\hat{A}_{\rm Ia}(\times 10^{-4})$ &  $1-p$   \\
MR01 & \textbf{0.410} &\textbf{2.51 $\pm${0.09}}& \textbf{0.99591}  & \textbf{0.221} &\textbf{2.18 $\pm${0.11}}& \textbf{0.99998}  & \textbf{0.478 }&\textbf{2.91 $\pm${0.14}}& \textbf{0.98690}  \\
MVP06 & \textbf{0.477} &\textbf{1.45 $\pm${0.05}}& \textbf{0.98716}  & \textbf{0.269} &\textbf{1.26} $\pm$\textbf{0.07}& \textbf{0.99990}  & 0.825 &1.73 $\pm${0.08}& 0.71272  \\
G05Wide & 1.364 &2.45 $\pm${0.09}& 0.10554  & 0.634 &2.12 $\pm${0.11}& 0.91908  & \textbf{0.259 }&\textbf{2.71 $\pm${0.13}}& \textbf{0.99993}  \\
G05Close & 0.899 &2.38 $\pm${0.07}& 0.60759  & \textbf{0.462} &\textbf{2.09 $\pm${0.10}}&\textbf{0.98979}  & 1.754 &3.19 $\pm${0.15}& 0.01124  \\
T08 & 0.674 &2.32 $\pm${0.08}& 0.88748  & \textbf{0.293} &\textbf{2.02 $\pm${0.11}}& \textbf{0.99977}  & \textbf{0.211} &\textbf{2.58 $\pm${0.13}}& \textbf{0.99999}  \\
P08 & 4.298 &2.18 $\pm${0.08}& 0.00000  & 2.523 &1.97 $\pm${0.10}& 0.00004  & 1.680 &2.42 $\pm${0.12}& 0.01802  \\
S05 & 2.148 &2.16 $\pm${0.08}& 0.00072  & 1.207 &1.94 $\pm${0.14}& 0.21784  & 2.440 &2.70 $\pm${0.17}& 0.00008  \\
Mix & 0.945 & 2.48$\pm${0.09} & 0.53928 & \textbf{0.449} & \textbf{2.15$\pm${0.11}} & \textbf{0.99058}  & \textbf{0.209} & \textbf{2.77$\pm${0.13}} & \textbf{0.99999}
\end{tblr}
\caption{Summary of the fit parameters found for all the tested combinations of DTDs and CSFRs. Each row corresponds to a DTD while each vertical block refers to a particular CSFR. Columns in each block contain the $\chi^2$ value per degree-of-freedom (left column), the optimal fraction $\hat{A}_{\rm Ia}$ (middle column) and the $1-p$ value of the $\chi^2$-test (right column). {For each CSFR, $p$ values lower than 0.05 (i.e, $1-p>$~0.95) are highlighted in bold}.}
\label{Table_chi2_fits_avg}
\end{table*}
\par In Figures \ref{Fig_RIafit_MD14_avg}, \ref{Fig_RIafit_H22_avg} and \ref{Fig_RIafit_JK23_avg} we show the final results, namely the theoretical cosmic Type Ia SN rates, as derived by adopting the CSFRs of Figure \ref{Fig_CSFR}, compared to SNIa data. In the left panels of these Figures, there is the evolution up to redshift $z$=2.5, while in the right panels the evolution is extended to a wider redshift range, from $z$=0 up to $z$=8. 
The cosmological time equivalent to the redshift is indicated on the upper axis, having assumed the standard $\Lambda$CDM scenario with $\Omega_\Lambda=0.70$, $\Omega_M=0.30$ and $H_0=70$~km s$^{-1}$~Mpc$^{-1}$ \citep{Hotokezaka19}. For the sake of simplicity, we have combined the literature values of $R_{\rm Ia}$ (see Table \ref{Table_binned_RIa}) in bins of similar redshift (black markers), weighting their contributions by the inverse of the errors. The resulting dispersion in $z$ is indicated by the horizontal errorbars, and is only significant within the range $1.0 \lesssim z \lesssim 1.5$. The reproduction of Figures \ref{Fig_RIafit_MD14_avg}, \ref{Fig_RIafit_H22_avg} and \ref{Fig_RIafit_JK23_avg} with the individual measurements of $R_{\rm Ia}$ is illustrated in Figures \ref{Fig_RIafit_MD14_ind}, \ref{Fig_RIafit_H22_ind} and \ref{Fig_RIafit_JK23_ind}, respectively.
%
\par In order to select the best DTD, we use a least-squares method to fit the data for each combination of CSFR and DTD, and then evaluate its quality by performing a $\chi^2$-test with $dof$ degrees of freedom, where $dof=N-1$ for the literature DTDs, $dof=N-2$ for the combination proposed in Sec. \ref{Sec_mixedDTDs} and $N$ is the number of observational data-points (see Table \ref{Table_chi2_fits_avg}). {The errors involved in the $\chi^2$ estimator are the quadratic sum of the observational errors and those computed for $R_{\rm IaCOSM}$ in Eq. \ref{Eq_R1aCosmo}. The latter depends on the free parameter we are trying to optimize ($A_{\rm Ia}$), introducing a non-linear dependence on this parameter and requiring an iterative approach to minimize $\chi^2$. Fortunately, this approach converges in few iterations}. For each combination of DTD and CSFR, we determine the fraction $\hat{A}_{\rm Ia}$ that minimises $\chi^2(A_{\rm Ia})$ (middle columns in Table \ref{Table_chi2_fits_avg}), in which we account for the asymmetry of the error-bars by using the upper (lower) errors when the fit over-estimates (sub-estimates) the observations. The error of $\hat{A}_{\rm Ia}$ is estimated from the width of the $\chi^2(A_{\rm Ia})$ curve near $\hat{A}_{\rm Ia}$ \citep{Fisher22}.
\subsection{Optimal combination of DTDs}
\label{Sec_mixedDTDs}
\begin{figure}
    \centering
    \includegraphics[width=0.98\linewidth]{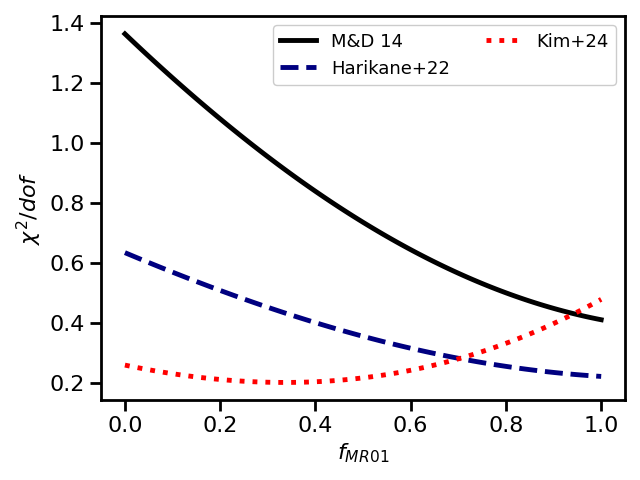}
    \caption{Goodness of fit of the real data $\chi^2$ per degree of freedom $(dof)$ as a function of the fraction of MR01 DTD ($f_{MR01}$) included in the combination with the G05Wide DTD. The lines correspond to the CSFR labelled in the legend.}
    \label{Fig_DTDcombination}
\end{figure}

Among the three theoretical DTDs considered in this work, only the G05Close DTD does not satisfactorily reproduce the observations of Type Ia SN rates, as noted in \citet{matteucci2009}. On the contrary, both the MR01 and the G05Wide DTDs proposed for the single and double degenerate scenarios, respectively, provide good fitting to the observations. Since these DTDs account for independent channels to produce Type Ia SNe, it is reasonable to estimate their individual contributions to the observed rates by testing multiple combinations of both. We parameterize such combinations by the fraction of MR01 DTD contribution, $f_{MR01}$, and repeat the fitting procedure of the observed data performed before. As Fig. \ref{Fig_DTDcombination} shows, an improvement of the fit is obtained only when Type Ia SN rates are computed assuming the \citet{jinkim23} CSFR, while for the \citet{madau14} and the \citet{harikane22} CSFR the MR01 alone leads to the optimal results. For the \citet{jinkim23} CSFR, the minimum of $\chi^2$ is {found at $f_{MR01}=0.34\pm0.01$ (i.e., 34\% of MR01 and 66\% of the G05Wide DTD)}, leading to a fit as good as that obtained with the empirical T08 DTD. This combined DTD is represented by the {black dash-dotted} lines in the corresponding figures.
{

%
%
%
%
%
\begin{figure*}
\begin{centering}
\includegraphics[width=0.98\textwidth]
{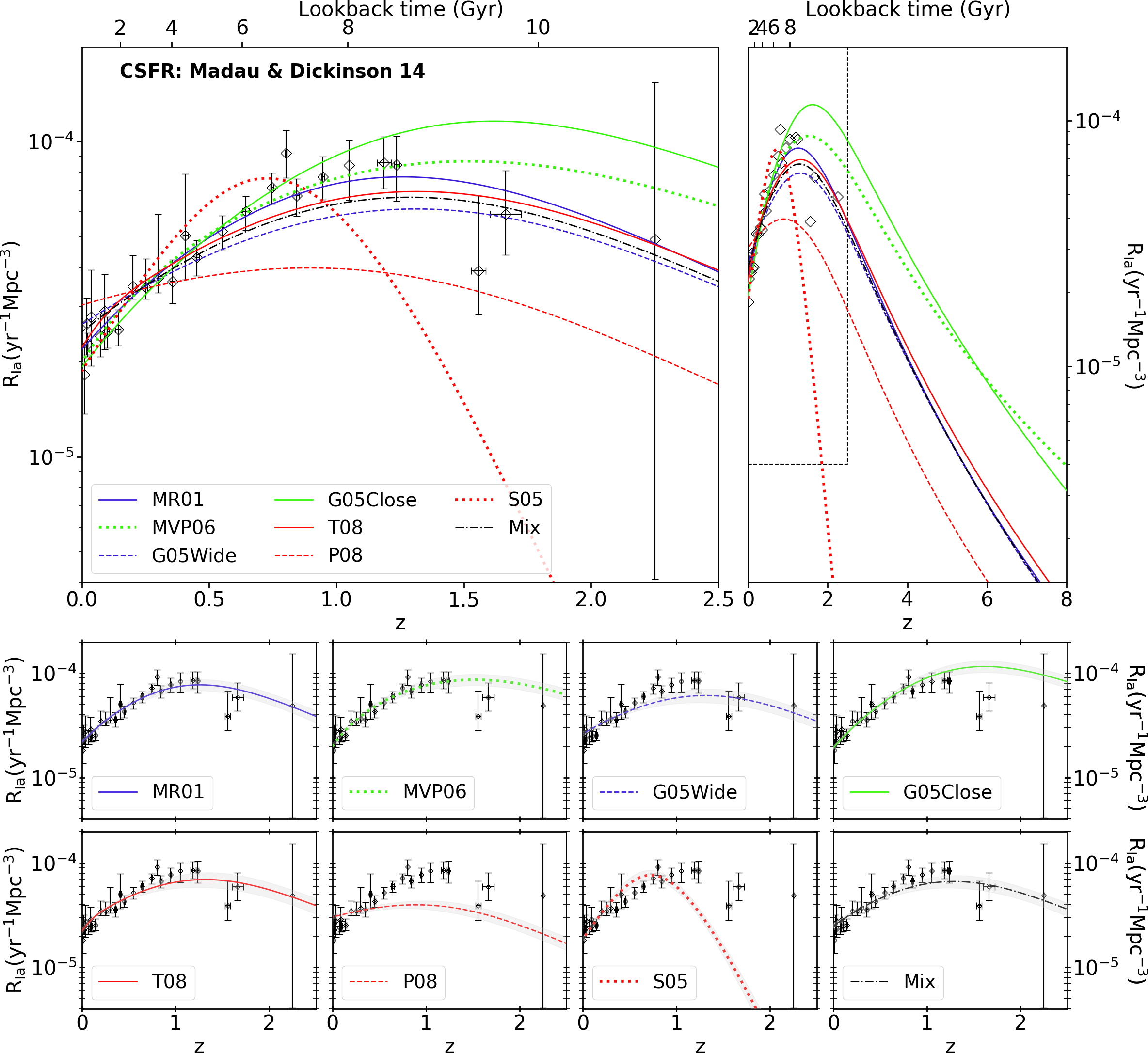}
\caption{Type Ia SN rate $R_{\rm Ia}$ as a function of redshift $z$ (lower horizontal axis) and age (upper horizontal axis) assuming the CSFR of \citep{madau14}. Black markers represent the binned observational data compiled from the literature and reported in Table \ref{Table_binned_RIa}. The curves correspond to the best fits assuming the DTDs used in \citet{palicioAS2023}, while the black dash-dotted line denotes a combination of the MR01 and G05Wide DTDs (see Section \ref{Sec_mixedDTDs}), both labelled in the legend. {Upper} right panel constitutes the zoomed-out version of the {upper} left one, enclosed by the {black dashed lines}. {Bottom panels illustrate the confidence intervals ($\pm 1 \sigma$) of the Type Ia SN rate computed from the errors in CSFR, assuming the DTDs indicated in each legend.}}
\label{Fig_RIafit_MD14_avg}
\end{centering}
\end{figure*}
%
%
%
%
%
%
\label{Sect_discussion}
 \par If we consider the data only up to $z=1.5$, which show smaller error bars, we can conclude that the best DTDs are those from MR01, T08 and G05Wide. In addition, we can completely exclude the S05 DTD, that assumes no prompt Type Ia SNe, and the G05Close DTDs as it underestimates (overestimates) the observations at redshifts lower (larger) than $z\approx 0.7$. This situation is reversed when the P08 DTD is considered.
\par The same conclusions about those DTDs were reached by \citet{matteucci2009}, who studied the effects of different Type Ia SN DTDs on the chemical evolution of the Milky Way, and concluded that the best DTDs are the SD and DD (Wide) ones. It should be noted that T08 is an empirical DTD and is similar to G05Wide, so we will not consider it in the following. 
\par As the left panels in Figs. \ref{Fig_RIafit_MD14_avg}, \ref{Fig_RIafit_H22_avg} and \ref{Fig_RIafit_JK23_avg} illustrate, the discrepancies among the Type Ia SN rates computed for different DTDs become more evident at high redshift, where the dependence on the cosmic SFR is also stronger. If we include in the comparison also this regime, in which the error bars of the observed Type Ia SN data are larger, then only the P08, S05 and G05Close DTDs can be firmly excluded.
\par These results confirm that the classical Type Ia SN scenarios (DD and SD) are the best to reproduce either the abundance patterns, for example the plot [$\alpha$/Fe] vs. [Fe/H] or the cosmic Type Ia SN rate. This means that the fraction of prompt Type Ia SNe {i.e. all the SNe exploding in the first 100 Myr since star formation} should not exceed 15-20\% of all Type Ia SNe, {as it is evident from Figure 1.} Moreover, as already pointed out in previous papers \citep[e.g.][]{matteucci2009, palicioAS2023} the differences between the SD scenario of MR01 and the wide DD scenario of G05 are negligible from the point of view of the cosmic rate as well as of the galactic chemical evolution. 
\par Concerning the adopted CSFRs, there is no much difference in the predicted cosmic Type Ia rates except for two cases: (i) when the CSFR of \citet{harikane22} is considered, the S05 DTD could be acceptable at low redshift, although it must be excluded anyway at high $z$ due to the large discrepancy with the data. (ii) Under the assumption of \citet{jinkim23} CSFR, the T08 provides the optimal fit of the observed Type Ia SN rate. {Although \citet{Hachisu2008} have proved the SD scenario can lead to such form of DTD}, the $t^{-1}$ trend of the T08 DTD has been historically considered supportive of the DD scenario \citep[see review of ][]{MaozMannucci12}. For instance, G05Wide and T08 have similar slope in Fig. \ref{Fig_DTD}. This connection is further reinforced by the improvement in data fit provided by the G05Wide DTD, proposed for the double degenerate scenario, when the \citet{jinkim23} CSFR is considered. Using a smoothed version of a T08-like DTD, \citet{Childress2014} were able to fit a subset of our observed Type Ia SN rates both assuming the \citet{Behroozi2013} CSFR and that inferred from their mass assembly model.
\par In contrast to T08 and G05Wide, the MVP06 DTD constitutes the second preferential delay time distribution when the single degenerate channel dominates; i.e, when the M\&D14 and H22 CSFRs are assumed. This clearly supports the need of two explosion regimes that account for the ``young'' and ``old'' progenitors, respectively \citep[the so-called ``A+B'' model]{Mannucci2005, Scannapieco2005}, especially for modelling the Type Ia SN rates observed in early type galaxies.
\par {We compared our results with existing literature. \citet{Graur14} explored two families of delay time distributions -- a power law and a Gaussian DTD -- to determine which one leads to the optimal fit for an observational cosmic Type Ia SN rate dataset, keeping the parameters of the mentioned DTD as free variables (i.e, the slope of the power law, the median and width of the Gaussian and the amplitudes of both). For all considered cosmic star formation rates, the resulting slopes of the power-law DTD were approximately -1, which made them compatible with the T08 DTD. However, the Gaussian DTD failed to reproduce the observations at a 95\% confidence level when the \citet{Behroozi2013} CSFR (their most similar CSFR to the M\&D14 CSFR tested in this work) was considered.}
\par {Additionally, \citet{Graur14} tested literature DTDs for both the single degenerate and double degenerate scenarios, and concluded the former leads to poor fits of the observed Type Ia SN rate. Even though a direct comparison with our results is not possible due to a different set of DTDs, we noted that their tests with the \citet{Yungelson10} DD DTD (similar to the T08 DTD) and \citet{Mennekens10} DD DTD (similar to our ``Mix'' DTD) provided visually satisfactory fits of their data-points (see their Figures 13 and 14). On the contrary, their DTDs for the SD scenario show very different late-regime slopes and minimum delay times compared to the MR01 DTD used in our work. This might explain why we obtained opposite conclusions regarding the single degenerate scenario.}
\par {More recently, \citet{Strolger20} proposed a family of DTDs based on the skew-normal distribution \citep{Ohagan76, Azzalini85, Azzalini05} to fit a compendium of Type Ia supernova rates (see their Table 5), while they considered a CSFR similar to that of \citet{madau14} within $\sim 1~\sigma$. As a result, their best fit suggested an exponential-like DTD that approximates the $t^{-1}$ T08 delay time distribution, typically associated to the double degenerate scenario (see discussion above).}
\par {Assuming the cosmic star formation rate from \citet{Behroozi2013}, \citet{Rodney14} tested a piecewise DTD characterised by two well-defined regimes: a plateau extending from a minimum delay of 40~Myr \citep{Belczynski05} to an abrupt transition at $t=0.5$~Gyr, followed by a $t^{-1}$ power law at later times ($t>0.5$~Gyr). The relative contribution of the prompt regime to the whole DTD is parameterised by the variable $f_p$. \citet{Rodney14} found $f_p$ ranges from 0.21 to 0.59 when considering only \textit{HST} \citep[CANDELS+CLASH][]{Grogin11, Koekemoer11, Postman12} and ground-based observations, respectively. Interestingly, the rough mean of these $f_p$ values provides a good approximation of our ``Mix'' DTD for $t\gtrsim0.2$~Gyr. When both \textit{HST} and ground-based observations are combined, they obtain a fraction $f_p\approx0.53$, similar to that found by \citet{mannucci2006}.}
\par Based on the results presented in this work, the single and double degenerate dichotomy is strongly dependent on the assumed CSFR. On one hand, those proposed by \citet{madau14} and \citet{harikane22} suggest a cosmic Type Ia SN rate dominated by the SD scenario, without invoking a degenerate companion. On the contrary, when considering the \citet{jinkim23} CSFR, we need at least a $\sim$~65\% of the total Type Ia SNe to be produced through the DD channel to explain the observations. However, as the authors point out, the CSFR proposed by \citet{jinkim23} flattens at $z\gtrsim 6$, diverging significantly from other CSFR models in the literature. {The obtained contribution of 34\% from SD progenitors to the total Type Ia SNe is of a similar order of magnitude to that proposed by \citet{Sternberg2011} for spirals (20-25\%), as well as to that inferred by \citet{Meng2009} for the SD white dwarf + main sequence star mechanism (30\%). However, it should be noted that the remaining 70\% might include other SD scenarios (e.g., WD+red giant, WD+helium star) as well as the DD channel}.
\section{Conclusions and Future perspectives}
\label{Sec_conclu}
\par In this work, we have evaluated multiple combinations of delay-time distributions and cosmic star formation rates to model a compilation of Type Ia SN rates reported in the literature. For the cosmic star formation rates, we have made use of the expressions proposed by \citet{madau14}, \citet{harikane22} and \citet{jinkim23} while for the DTDs we considered those explored in \citet{palicioAS2023}, including the theoretical formulations for the single and double degenerate channels from \citet{matteucci2001} and \citet{greggio2005}, respectively, as well as the empirical ones from other works. Using a least-square fitting technique, we determined the optimal fraction of Type Ia SN progenitors, $\hat{A}_{\rm Ia}$, for each combination of CSFRs and DTDs, whose goodness of fit is quantified by the p-value of the associated $\chi^2$ test.
We found that, when the \citet{madau14} and \citet{harikane22} CSFRs are considered, the MR01 DTD proposed for the single degenerate scenario provides the best modelling for the observed Type Ia SN rates. For the case of the \citet{jinkim23} CSFR, however, the empirical delay-time distribution suggested by \citet{totani2008} leads to the optimal fit of the observations. Though not optimal, the G05Wide DTD proposed by \citet{greggio2005} constitutes a good alternative to the T08 DTD for the latter CSFR. If we relax the criteria for a good fit to $1-p > 0.95$ (values highlighted in bold in Table \ref{Table_chi2_fits_avg}), then only the MR01 DTD provides satisfactory fits for all the CSFRs (also the MVP06 DTD if we consider the results in Table \ref{Table_chi2_fits_ind}). However, for the H22 and K24 CSFRs, we cannot resolve the SD/DD degeneracy.
\par Finally, by combining the theoretic DTDs proposed for the single and double degenerate scenarios, we found the MR01 DTD alone is the best to reproduce the observed Type Ia SN rates in two of the three tested CSFRs (\citealt{madau14} and \citealt{harikane22}), while it requires 70\% of the G05Wide to provide better results when the \citet{jinkim23} CSFR is considered. This dichotomy in the best scenario determination seems to be motivated by the different trend of the \citet{jinkim23} CSFR at high $z$, when compared to the other CSFRs. The observations at higher redshift allowed us to break the degeneracy noted by \citet{bonaparte2013} in the determination of the dominant channel for the Type Ia SNe, giving preference to the single degenerate scenario, although been strongly dependent on the CSFR assumption.
 \par In conclusion, our results prove that, even in the most favourable scenario for the double degenerate channel, at least the $\sim$ 34\% of the observed Type Ia SNe must be produced by a white dwarf with a non-degenerate companion. This implies the single degenerate channel constitutes a viable scenario that should not be obviated \citep{Livio2000}, offering an alternative mechanism to the sub-Chandra Type Ia-producing mergers, which were proposed to account for the limited number of close WD binaries capable of producing Type Ia SNe within the Hubble time \citep{Maoz2008, Mennekens10}. 
 However, our results show clearly that other proposed Type Ia SN scenarios do not reproduce the cosmic Type Ia SN rate, irrespective of the assumed CSFR,  nor the [$\alpha$/Fe] vs. [Fe/H] relation in the Milky Way \citep{matteucci2009}.
 
 In the future, JWST \citep{Gardner06, Rigby23} will hopefully provide more data of SNe Ia at high $z$, allowing us to anchor the predicted SN rates at earlier times, where the effect of the adopted DTD is more evident.

}

\section*{Acknowledgement}
We thank the anonymous referee whose comments and corrections improved this manuscript. P. A. Palicio acknowledges the financial support from the Centre national d’études spatiales (CNES).
This work was partially supported by the European Union (ChETEC-INFRA, project no. 101008324). F. Matteucci thanks I.N.A.F. for the 1.05.12.06.05 Theory Grant - Galactic archaeology with radioactive and stable nuclei.
\bibliographystyle{aa} 
\bibliography{biblio}
\FloatBarrier
\begin{appendix}
\onecolumn
\section{Additional figures}
%
%
%
%
%
%
%
\begin{figure*}[!ht]
\begin{centering}
\includegraphics[width=0.95\textwidth]{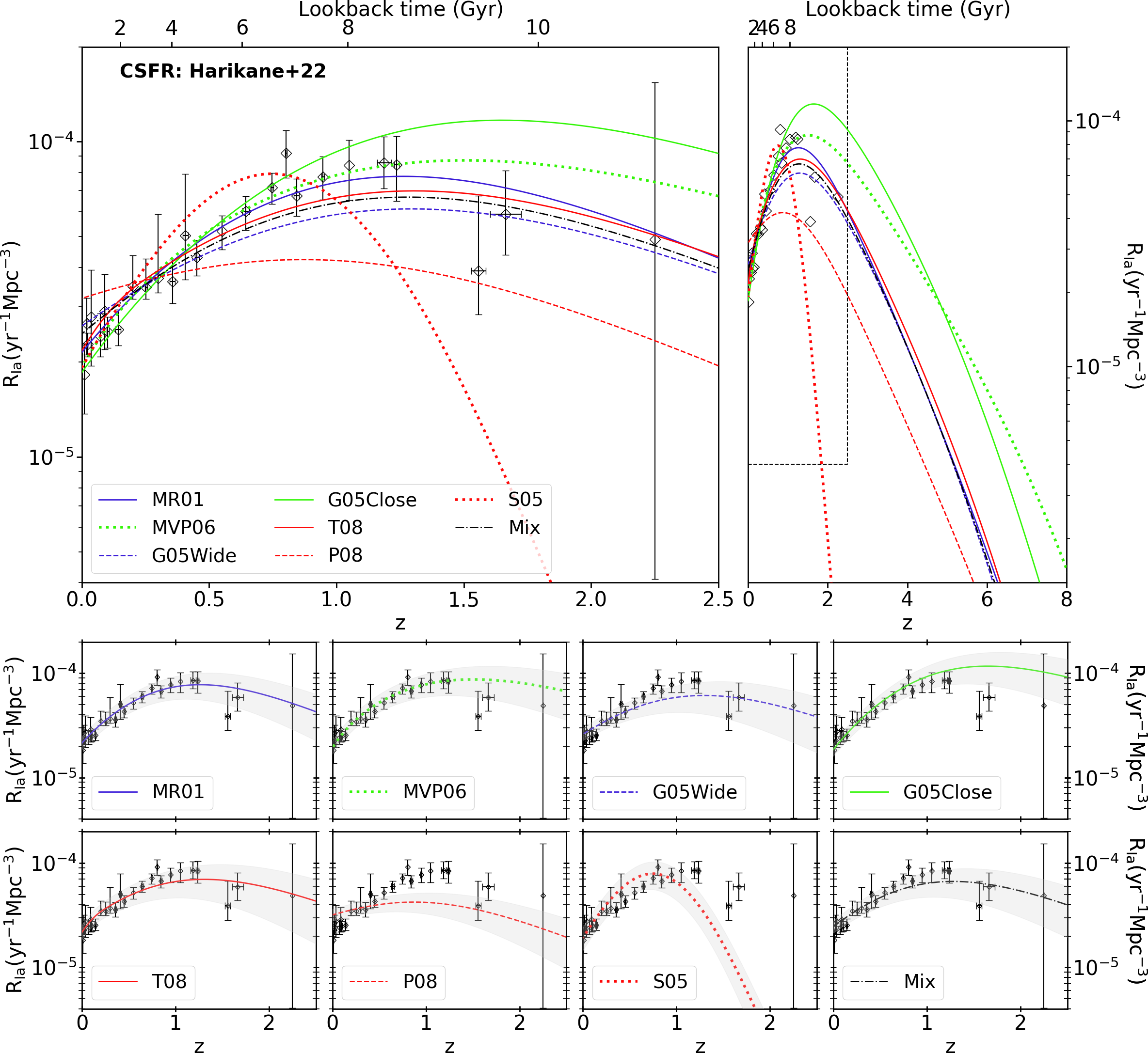}
\caption{Same as Fig. \ref{Fig_RIafit_MD14_avg} but for the CSFR of \citet{harikane22}}
\label{Fig_RIafit_H22_avg}
\end{centering}
\end{figure*}
\begin{figure*}[!h]
\begin{centering}
\includegraphics[width=0.95\textwidth]{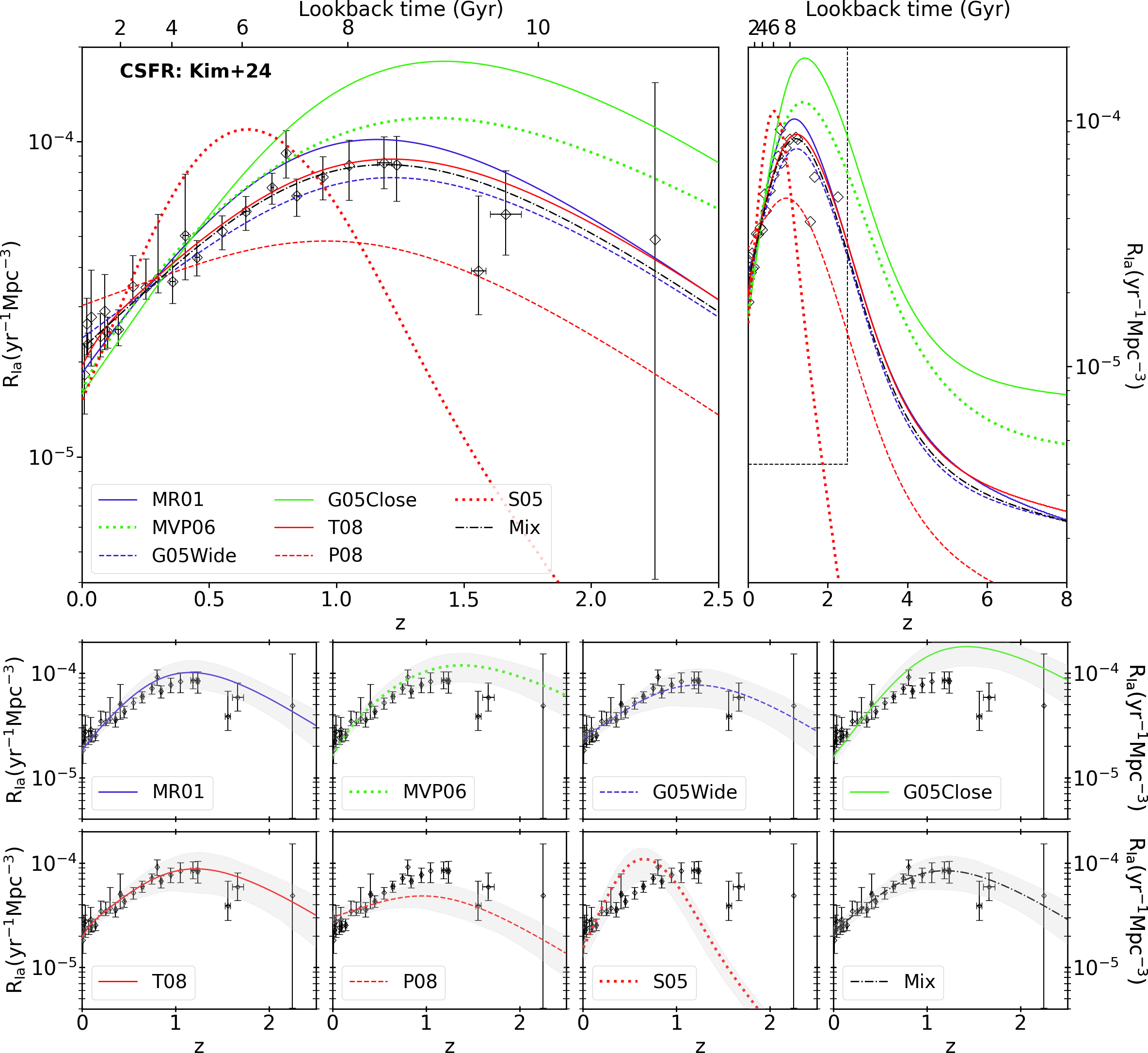}
\caption{Same as Fig. \ref{Fig_RIafit_MD14_avg} but for the CSFR of \citet{jinkim23}.}
\label{Fig_RIafit_JK23_avg}
\end{centering}
\end{figure*}
\FloatBarrier
\section{Table of individual Type Ia SN rates}
\par Table with the individual measurements binned and averaged in Table \ref{Table_binned_RIa}.\nopagebreak
\begin{longtable}{ccccc}
\caption{Compilation of individual measurements of observed Type Ia SN rates at different redshifts used in this work.}\\
\label{Table_full_RIa}\\
\hline
\hline
  z	&	       Rate	&	$\rm \sigma +$	&	$\rm \sigma -$& Reference\\ 
\hline
\endfirsthead
\caption{continued}\\
\hline
  z	&	       Rate	&	$\rm \sigma +$	&	$\rm \sigma -$& Reference\\ 
\hline
\endhead
\hline
\endfoot
\hline
\endlastfoot
0.01	&	0.183	&	+0.046	&	-0.046	&	\citet{Cappellaro99} \\
0.02	&	0.265	&	+0.055	&	-0.054	&	\citet{li2010} \\
{0.024}	&	{0.228}	&	{+0.020}	&	{-0.020}	&	{\citet{Desai2024}} \\
0.0375	&	0.278	&	+0.115	&	-0.083	&	\citet{Dilday10} \\
0.073	&	0.243	&	+0.043	&	-0.034	&	\citet{Frohmaier19} \\
0.09	&	0.29	&	+0.09	&	-0.07	&	\citet{Dilday08} \\
0.098	&	0.24	&	+0.12	&	-0.12	&	\citet{Madgwick2003} \\
0.1	&	0.55	&	+0.538	&	-0.352	&	\citet{Cappellaro15} \\
0.1		&0.259	&	+0.055	&	-0.044	&	\citet{Dilday10} \\
0.11	&	0.247	&	+0.034	&	-0.040	&	\citet{Graur13} \\
0.13	&	0.158	&	+0.066	&	-0.055	&	\citet{Blanc04} \\
0.14	&	0.28	&	+0.23	&	-0.136	&	\citet{Hardin00} \\
0.15	&	0.307	&	+0.051	&	-0.034	&	\citet{Dilday10} \\
0.15	&	0.32	&	+0.240	&	-0.237	&	\citet{RodneyTonry10}\\
0.2	 &	0.348	&	+0.088	&	-0.031	&	\citet{Dilday10} \\
0.25	&	0.365	&	+0.184	&	-0.030	&	\citet{Dilday10} \\
0.25	&	0.36	&	+0.611	&	-0.436	&	\citet{Rodney14} \\
0.25	&	0.39	&	+0.164	&	-0.156	&	\citet{Cappellaro15} \\
0.26	&	0.32	&	+0.106	&	-0.113	&	\citet{Perrett12} \\
0.3	 &	0.34	&	+0.264	&	-0.266	&	\citet{Botticella08}\\
0.3	 &	0.434	&	+0.397	&	-0.0375	 & \citet{Dilday10} \\
0.35	& 	0.34	&	+0.202	&	-0.192	&	\citet{RodneyTonry10} \\	
0.35	&	0.41	&	+0.09	&	-0.098	&	\citet{Perrett12} \\
0.368	&	0.31	&	+0.094	&	-0.058	&	\citet{Neill07} \\
0.40	&	0.69	&	+1.577	&	-0.367	&	\citet{Dahlen04} \\
0.40	&	0.53	&	+0.39	&	-0.17	&	\citet{Kuznetsova08} \\
0.42	&	0.46	&	+0.43	&	-0.345	&	\citet{Graur14} \\
0.44	&	0.262	&	+0.236	&	-0.179	&	\citet{Okumura14} \\
0.45	&	0.73	&	+0.24	&	-0.24	&	\citet{BarrisTonry06} \\
0.45	&	0.31	&	+0.192	&	-0.155	&	\citet{RodneyTonry10} \\
0.45	&	0.41	&	+0.086	&	-0.092	&	\citet{Perrett12} \\
0.45	&	0.52	&	+0.194	&	-0.206	&	\citet{Cappellaro15} \\
0.46	&	0.48	&	+0.17	&	-0.17	&	\citet{Tonry03} \\
0.467	&	0.42	&	+0.143	&	-0.108	&	\citet{Neill06} \\
0.47	&	0.8	&	+1.700	&	-0.375	&	\citet{Dahlen08} \\
0.55	&	0.568	&	+0.138	&	-0.124	&	\citet{Pain02} \\
0.55	&	0.32	&	+0.156	&	-0.156	&	\citet{RodneyTonry10} \\
0.55	&	0.55	&	+0.086	&	-0.092	&	\citet{Perrett12} \\
0.552	&	0.63	&	+0.278	&	-0.287	&	\citet{Neill07} \\
0.62	&	1.29	&	+0.920	&	-0.635	&	\citet{Melinder12} \\
0.65	&	1.49	&	+0.31	&	-0.31	&	\citet{BarrisTonry06} \\
0.65	&	0.49	&	+0.220	&	-0.187	&	\citet{RodneyTonry10} \\
0.65	&	0.55	&	+0.078	&	-0.092	&	\citet{Perrett12} \\
0.65	&	0.69	&	+0.330	&	-0.324	&	\citet{Cappellaro15} \\
0.714	&	1.13	&	+0.572	&	-0.725	&	\citet{Neill07} \\
0.74	&	0.43	&	+0.36	&	-0.32	&	\citet{Poznanski2007b} \\
0.74	&	0.79	&	+0.33	&	-0.41	&	\citet{Graur11} \\
0.75	&	1.78	&	+0.34	&	-0.34	&	\citet{BarrisTonry06} \\
0.75	&	0.68	&	+0.311	&	-0.252	&	\citet{RodneyTonry10} \\
0.75	&	0.67	&	+0.092	&	-0.106	&	\citet{Perrett12} \\
0.75	&	0.51	&	+0.354	&	-0.268	&	\citet{Rodney14} \\
0.80	&	1.57	&	+0.869	&	-0.586	&	\citet{Dahlen04} \\
0.80	&	0.93	&	+0.25	&	-0.25	&	\citet{Kuznetsova08} \\
0.80	&	0.839	&	+0.237	&	-0.220	&	\citet{Okumura14} \\
0.807	&	1.18	&	+0.744	&	-0.53	&	\citet{Barbary12} \\
0.83	&	1.30	&	+0.801	&	-0.577	&	\citet{Dahlen08} \\
0.85	&	0.78	&	+0.380	&	-0.272	&	\citet{RodneyTonry10} \\
0.85	&	0.66	&	+0.092	&	-0.1	&	\citet{Perrett12} \\  
0.94	&	0.45	&	+0.255	&	-0.199	&	\citet{Graur14} \\
0.95	&	0.76	&	+0.406	&	-0.360	&	\citet{RodneyTonry10} \\
0.95	&	0.89	&	+0.15	&	-0.166	&	\citet{Perrett12} \\     
1.05	&	0.79	&	+0.45	&	-0.496	&	\citet{RodneyTonry10} \\
1.05	&	0.85	&	+0.184	&	-0.205	&	\citet{Perrett12} \\    
1.14	&	0.705	&	+0.259	&	-0.209	&	\citet{Okumura14} \\
1.187	&	1.33	&	+0.947	&	-0.55	&	\citet{Barbary12} \\
1.20	&	1.15	&	+0.568	&	-0.511	&	\citet{Dahlen04} \\
1.20	&	0.75	&	+0.35	&	-0.30	&	\citet{Kuznetsova08} \\
1.21	&	1.32	&	+0.523	&	-0.431	&	\citet{Dahlen08} \\
1.23	&	1.05	&	+0.45	&	-0.56	&	\citet{Poznanski2007b} \\
1.23	&	0.84	&	+0.25	&	-0.28	&	\citet{Graur11} \\
1.25	&	0.64	&	+0.460	&	-0.318	&	\citet{Rodney14} \\
1.535	&	0.77	&	+1.156	&	-0.94	&	\citet{Barbary12} \\
1.55	&	0.12	&	+0.58	&	-0.12	&	\citet{Kuznetsova08} \\
1.59	&	0.45	&	+0.343	&	-0.237	&	\citet{Graur14} \\
1.60	&	0.44	&	+0.349	&	-0.273	&	\citet{Dahlen04} \\
1.61	&	0.42	&	+0.433	&	-0.269	&	\citet{Dahlen08} \\
1.67	&	0.81	&	+0.79	&	-0.60	&	\citet{Poznanski2007b} \\
1.69	&	1.02	&	+0.54	&	-0.37	&	\citet{Graur11} \\
1.75	&	0.72	&	+0.672	&	-0.410	&	\citet{Rodney14} \\
2.1		& <1.07	&	0.0	&	0.0	&	\citet{Graur14} \\
2.25	&	0.49	&	+1.05	&	-0.449	&	\citet{Rodney14} \\
\hline
\multicolumn{5}{l}{\small Individual measurements of observed Type Ia SN rates (second column)}\\
\multicolumn{5}{l}{\small at different redshifts (first column) including the upper and lower errors}\\
\multicolumn{5}{l}{\small (third and fourth columns, respectively). The rates and their errors are}\\
\multicolumn{5}{l}{\small given in units of $10^{-4}$yr$^{-1}$Mpc$^{-3}$. References can be found in the last}\\
\multicolumn{5}{l}{\small column.} \\
\end{longtable}
%
%
%
%
\section{Results with the individual measurements}
\begin{table*}[!h]
\centering
\begin{tblr}{
  cell{1}{2} = {c=3}{c},
  cell{1}{5} = {c=3}{c},
  cell{1}{8} = {c=3}{c},
  cell{2}{2} = {c},
  cell{2}{3} = {c},
  cell{2}{4} = {c},
  cell{2}{5} = {c},
  cell{2}{6} = {c},
  cell{2}{7} = {c},
  cell{2}{8} = {c},
  cell{2}{9} = {c},
  cell{2}{10} = {c},
  vline{3,6,9} = {1}{},
  vline{1-2,5,8,11} = {2-10}{},
  hline{1} = {2-10}{},
  hline{2-3,10} = {-}{},
  hline{2-3,11} = {-}{},
}
         &  M\&D14 CSFR       &          &       & H22 CSFR         &          &       & K24 CSFR         &          &       \\
DTD      & $\chi^2/ dof$ & $\hat{A}_{\rm Ia} (\times 10^{-4}) $   & $1-p$   & $\chi^2/ dof$ & $\hat{A}_{\rm Ia}(\times 10^{-4})$   & $1-p$   & $\chi^2/ dof$ & $\hat{A}_{\rm Ia}(\times 10^{-4})$   & $1-p$   \\
MR01 & \textbf{0.692} & \textbf{2.57} $\pm$ \textbf{0.08} & \textbf{0.983} &\textbf{0.549} & \textbf{2.23$\pm${0.09}} & \textbf{0.999}  &\textbf{ 0.593} & \textbf{2.92$\pm${0.11}} & \textbf{0.998}  \\
MVP06 & \textbf{0.700 }& \textbf{1.47$\pm${0.04}} & \textbf{0.981}  &\textbf{0.552 }& \textbf{1.28$\pm${0.05} }& \textbf{0.999}  & \textbf{0.736} & \textbf{1.71$\pm${0.07}} & \textbf{0.962}  \\
G05Wide & 1.053 & $2.52\pm{0.08}$ & 0.353  & 0.769 & $2.22\pm{0.10}$ & 0.936  & \textbf{0.577} & \textbf{2.81}$\pm$\textbf{0.11} &\textbf{0.999}  \\
G05Close & 0.876 & $2.43\pm{0.07}$ & 0.777  &\textbf{0.627 }& \textbf{2.10$\pm${0.09}} & \textbf{0.996}  & 1.285 & $3.00\pm{0.12}$ & 0.045  \\
T08 & 0.792 & $2.37\pm{0.08}$ & 0.912  & \textbf{0.590} & \textbf{2.08$\pm${0.10}} & \textbf{0.998}  & \textbf{0.500 }& \textbf{2.63}$\pm$\textbf{{0.11}} & \textbf{0.999}  \\
P08 & 2.026 & $2.22\pm{0.07}$ & 0.000  & 1.521 & $2.04\pm{0.09}$ & 0.002  & 1.211 & $2.52\pm{0.11}$ & 0.097  \\
S05 & 1.186 & $2.12\pm{0.07}$ & 0.124  & 0.829 & $1.86\pm{0.10}$ & 0.861  &1.361 & $2.33\pm{0.12}$ & 0.018  \\
Mix & 0.896 & 2.54$\pm${0.08} & 0.731 & \textbf{0.679} & \textbf{2.22}$\pm$\textbf{0.09} & \textbf{0.987  }&\textbf{ 0.534} & \textbf{2.85$\pm${0.11}} & \textbf{0.999}
\end{tblr}
\caption{Summary of the fit parameters found for all the tested combinations of DTDs and CSFRs, considering the individual measurements of $R_{\rm Ia}(z)$ of Table \ref{Table_full_RIa}. Each row corresponds to a DTD while each vertical block refers to a particular CSFR. Columns in each block contain the $\chi^2$ value per degree-of-freedom (left column), the optimal fraction $\hat{A}_{\rm Ia}$ (middle column) and the $1-p$ value of the $\chi^2$-test (right column). {For each CSFR, $p$ values lower than 0.05 (i.e, $1-p>$~0.95) are highlighted in bold.} }
\label{Table_chi2_fits_ind}
\end{table*}
\begin{figure*}
\begin{centering}
\includegraphics[width=0.98\textwidth]{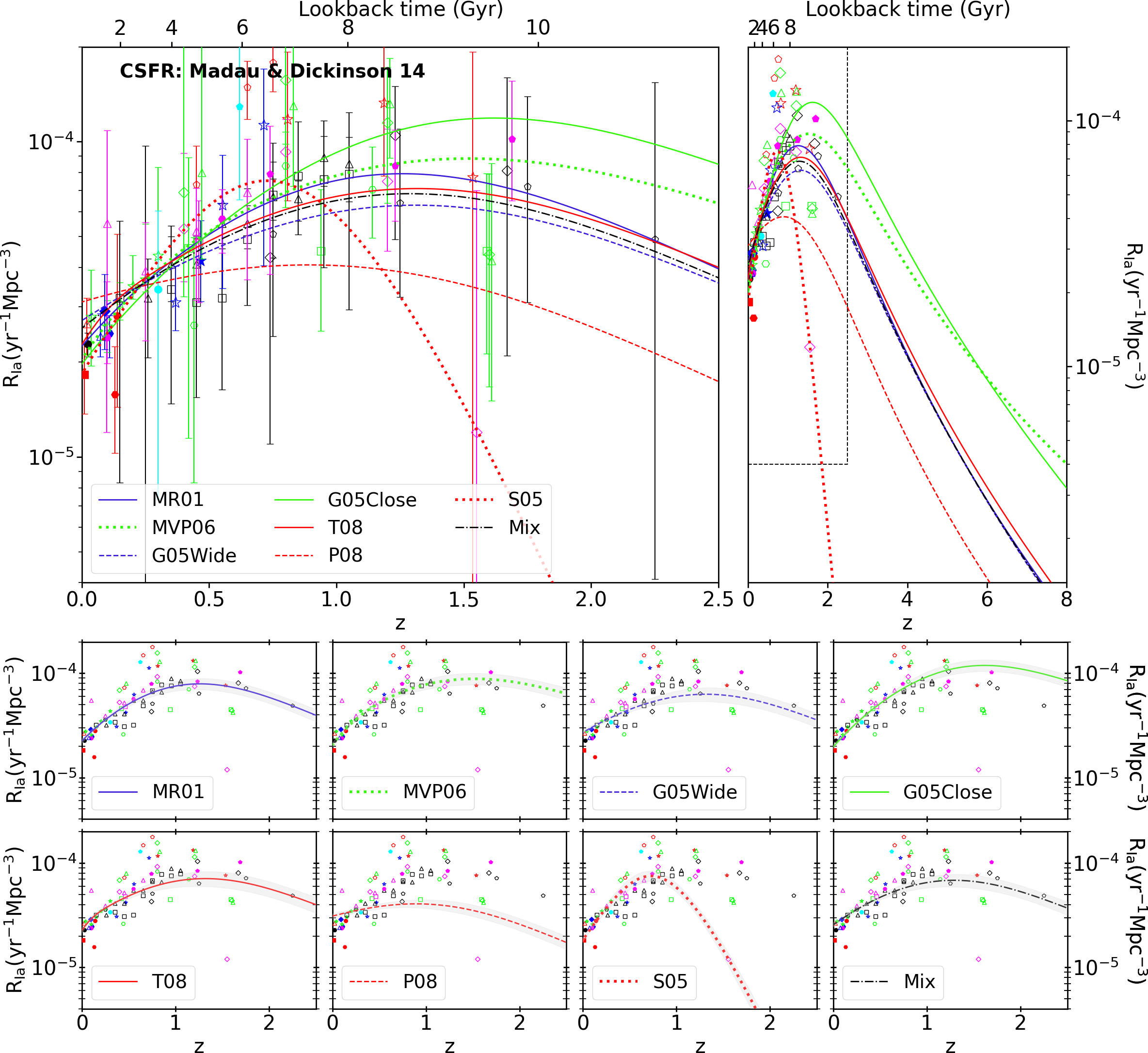}
\caption{Type Ia SN rate $R_{\rm Ia}$ as a function of redshift $z$ (lower horizontal axis) and age (upper horizontal axis) assuming the CSFR of \citep{madau14}. Each marker-colour pair corresponds to one reference in literature: \citealt{li2010} (open red triangles), \citealt{Neill06} (filled blue stars), \citealt{BarrisTonry06} (open red pentagons), \citealt{Graur13} (filled blue hexagons), \citealt{Madgwick2003} (filled magenta circles), \citealt{Poznanski2007b} (open black diamonds), \citealt{Cappellaro99} (filled red squares), \citealt{Frohmaier19} (open blue triangles), \citealt{Barbary12} (open red stars), \citealt{Melinder12} (filled cyan pentagons), \citealt{Blanc04} (filled red hexagons), \citealt{Hardin00} (filled red circles), \citealt{Kuznetsova08} (open magenta diamonds), \citealt{Graur14} (open lime squares), \citealt{Perrett12} (open black triangles), \citealt{Dilday10} (open lime stars), \citealt{Rodney14} (open black pentagons), \citealt{Okumura14} (open lime hexagons), \citealt{Botticella08} (filled cyan circles), \citealt{Dahlen04} (open lime diamonds), \citealt{RodneyTonry10} (open black squares), \citealt{Cappellaro15} (open magenta triangles), \citealt{Neill07} (open blue stars), \citealt{Graur11} (filled magenta pentagons), \citealt{Pain02} (filled magenta hexagons), \citealt{Dilday08} (filled blue diamonds), \citealt{Tonry03} (open magenta squares), \citealt{Dahlen08} (open lime triangles) {and \citealt{Desai2024} (solid black circle)}. The curves represent the best fits for the assumed DTDs (labelled in the legend). {Upper} right panel constitutes the zoomed-out version of the {upper} left one (enclosed by the dashed lines) without showing the errorbars for sake of visualisation. {Bottom panels illustrate the confidence intervals ($\pm 1 \sigma$) of the Type Ia SN rate computed from the errors in CSFR, assuming the DTDs indicated in each legend.}}
\label{Fig_RIafit_MD14_ind}
\end{centering}
\end{figure*}

\begin{figure*}
\begin{centering}
\includegraphics[width=0.98\textwidth]
{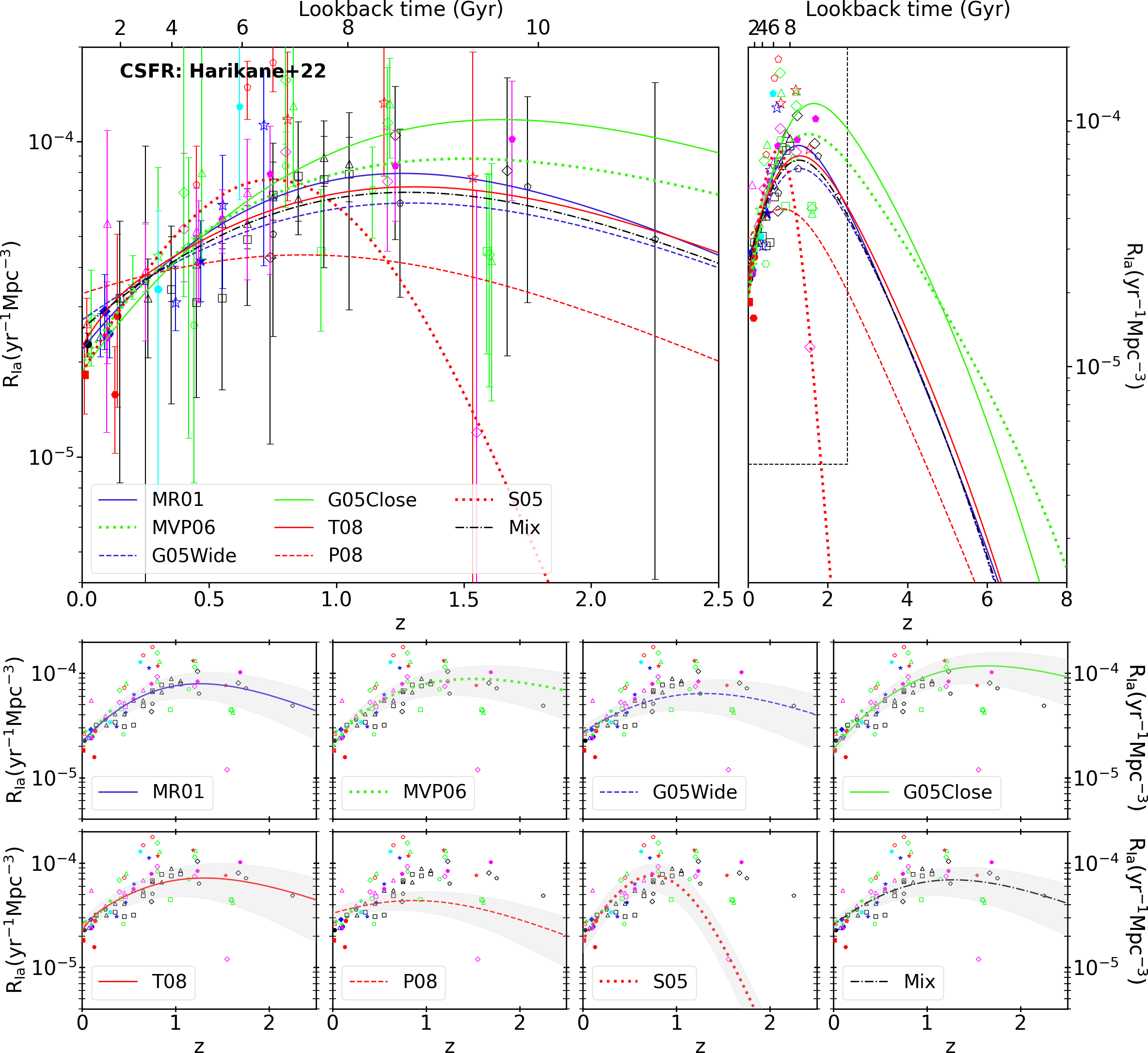}
\caption{Same as Fig. \ref{Fig_RIafit_MD14_ind} but for the CSFR of \citet{harikane22}}
\label{Fig_RIafit_H22_ind}
\end{centering}
\end{figure*}

\begin{figure*}
\begin{centering}
\includegraphics[width=0.98\textwidth]
{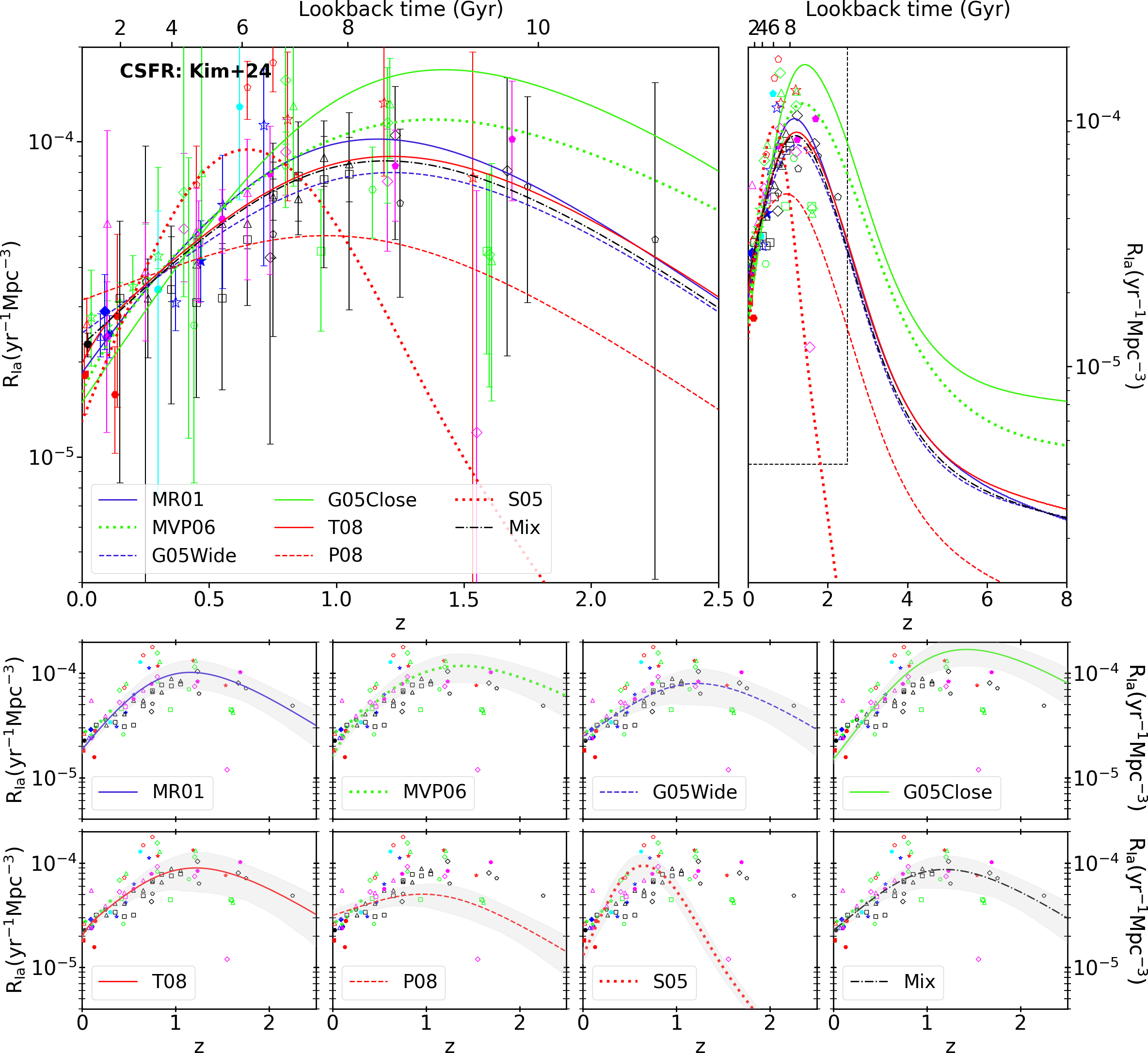}
\caption{Same as Fig. \ref{Fig_RIafit_MD14_ind} but for the CSFR of \citet{jinkim23}}
\label{Fig_RIafit_JK23_ind}
\end{centering}
\end{figure*}
\end{appendix}
\end{document}